\documentclass[seceq]{ptptex}
\usepackage{wrapft}

\usepackage{amsmath}
\usepackage{comment}
\newcommand{\half}{ \frac{1}{2} }
\newcommand{\dotX}{\dot{X}}
\newcommand{\dotY}{\dot{Y}}

\newcommand{\ket}[1]{\big|~#1~\big\rangle}

\newcommand{\Exp}[1]{\big\langle~#1~\big\rangle}

\title{
Charged Strings and Spectrum-Generating Algebra
}

\author{
Akira \textsc{Kokado}$^{1,}$\footnote{E-mail: kokado@kobe-kiu.ac.jp},  
Gaku \textsc{Konisi}$^{2,}$\footnote{E-mail: konisi@womat.zaq.ne.jp} 
and Takesi \textsc{Saito}$^{2,}$\footnote{E-mail: tsaito@k7.dion.ne.jp}
}

\inst{
$^1$Kobe International University, Kobe 658-0032, Japan\\
$^2$Department of Physics, Kwansei Gakuin University, Sanda, 669-1337, Japan
}

\abst{
We consider a system that both ends of a charged open string are attached on the D$p$-brane 
with constant electromagnetic fields. Contrary to neutral strings, the quantization of 
charged strings has not been so far considered well.  For this system we construct the 
spectrum-generating algebra (SGA), which involves the cyclotron frequency. 
When the cyclotron frequency is set to be zero, the SGA is reduced to the ordinary SGA for neutral strings. 
The new SGA for charged strings guarantees that this system is ghost-free if certain conditions are satisfied. 
We also consider its application to the Hall effect for charged strings.}

\begin{document}

\maketitle
\section{Introduction}
The charged open string is defined as $X_\mu (\tau ,\sigma )$,  $0\leq \sigma \leq \pi $, 
which ends on the D$p$-brane with different charges $q_0$ at $\sigma =0$ and  $q_\pi $ at 
$\sigma =\pi $, whereas $q_0 + q_\pi =0$ for the neutral string. The D$p$-brane involves 
a constant background field perpendicular to the brane. 

Contrary to the neutral string, the quantization of the charged string is not so well
 known.\cite{rf:Abouelsaood, rf:Chu, rf:kokado2000} Here we meet a problem of how to work the no-ghost theorem 
in this system. For the neutral string with background fields, the spectrum-generating algebra 
(SGA) is well known to guarantee the system to be ghost-free, if the space-time dimension is $d=26$ 
and the Regge intercept is $\alpha (0)=1$. For the charged string with a background field, however, 
it is not so clear whether the conventional SGA works well or not, because of existence of normal 
modes involving the cyclotron frequency. We then construct the new SGA for charged strings, 
and find that our system is ghost-free if $d=26$ and $\alpha (0)=1-\sum_i\omega_i /2 + \sum_i{\omega_i}^2/2$, 
where $\omega _i$ is the $i$-th block cyclotron frequency of the charged string. 

As an application, we consider the Hall effect for charged strings on the D2-brane, in which constant 
electromagnetic fields are living. We calculate the Hall conductivity for the current as a flow of charged strings.

In Sec.\ref{sec:qunantize} we summarize the quantization of the charged string placed in constant background fields. 
In Sec.\ref{sec:SGAstring} the SGA for charged strings is constructed. In Sec.\ref{sec:Hall_cond} we calculate the Hall conductivity 
for the charged strings. The final section is devoted to concluding remarks.


\section{Quantization of the charged string} \label{sec:qunantize}
The interaction Lagrangian of our system is given by
\begin{align}
  & L_I
  = q_\sigma \, \dotX_\mu(\tau, \sigma)\, A^\mu\big( X(\tau, \sigma) \big)~
      \Big\vert_{\sigma=0}
   +q_\sigma \, \dotX_\mu(\tau, \sigma)\, A^\mu\big( X(\tau, \sigma) \big)~
      \Big\vert_{\sigma=\pi }. 
\label{eq:def-L_int}
\end{align}
where $A^\mu $, $\mu =0, 1, \cdots, p$ is the $U(1)$ gauge field living on the D$p$-brane.

we choose a gauge such as
\begin{align}
  & A^\mu = -\frac{1}{2}F^{\mu}_{\ \nu}~X^\nu~,
\label{eq:gauge}
\end{align}
with $F^{\mu}_{\ \nu}$, the constant electromagnetic field strength. Introducing a 
time-indepent function $\rho (\sigma )$ such that
\begin{align}
  & \rho(0) = q_0~, & & \rho(\pi) = -q_\pi ~,
\end{align}
the interaction Lagrangian can be rewritten as
\begin{align}
   L_I
  &= \half~\Big[~\rho(\sigma)\, \dotX_\mu(\tau, \sigma)\,
     F^{\mu}_{\ \nu}\, X^\nu(\tau, \sigma)~\Big]^\pi_0
  = \half \int^\pi_0 d\sigma~\partial_\sigma \Big(
     \rho(\sigma)\, \dotX_\mu(\tau, \sigma)\,
        F^{\mu}_{\ \nu}\, X^\nu(\tau, \sigma) \Big)
\nonumber \\
  &= \half \int^\pi_0 d\sigma~\Big(
     \rho'\, \dotX_\mu\, F^{\mu}_{\ \nu}\, X^\nu
  + 2\, \rho\, \dotX_\mu\, F^{\mu}_{\ \nu}\, \big( X^\nu \big)' \Big)~,
\label{eq:def-L_int-II}
\end{align}
where the total time derivative $\partial_\tau\big( \rho X' F X \big)$
has been discarded.
This theory does not depend on the functional form of $\rho(\sigma)$
for $0 < \sigma < \pi$,
because the action based on Eq.(\ref{eq:def-L_int-II})
is invariant under a variation with respect to $\rho(\sigma)$.

The total Lagrangian is given as
\begin{align}
  & L
  = \half \int^\pi_0 d\sigma~
       \left[~\dotX_\mu\, \dotX^\mu
    - {X_\mu}' {X^\mu}'~\right]
  + \half \int^\pi_0 d\sigma~\dotX_\mu F^{\mu}_{\ \nu} \left[~
     \rho'\, X^\nu + 2 \rho\,  {X^\nu}'~\right]~.
\label{eq:def-L}
\end{align}
The equation of motion and the boundary conditions
follow from the action based on Eq.(\ref{eq:def-L}):
\begin{align}
  & \ddot{X}^\mu - \big( X^\mu \big)'' = 0~, \qquad  \mu = 0,1, \cdots , d-1~, 
\label{eq:EOM_X} \\
  & \left[~\big( X^\mu \big)'
  + \rho F^\mu{}_\nu \dotX^\nu~\right]\,  \Big\vert_{\sigma=0, \pi} = 0 \qquad  \mu ,\nu = 0,1, \cdots , p ~, 
\label{eq:bc_X} \\
  & X^a(\tau ,\sigma )\Big\vert_{\sigma=0} = c^a_0~, \ \  X^a(\tau ,\sigma )\Big\vert_{\sigma=\pi } = c^a_\pi~,
 \ \ a=p+1, \cdots, d-1~,
\label{eq:bc_Xa} 
\end{align}
where $c_0^a, c_\pi ^a$ are constants, indicating locations of D-branes.

We will concentrate on a $2\times 2$ block of $F^\mu _{\ \nu }$
\begin{align}
  & F^{\mu}_{\ \nu}
  = \begin{pmatrix} 0 &  K \\ -K & 0 
    \end{pmatrix}~.
\label{eq:def-2F2}
\end{align}
where $K$ is a positive real number. This means that we are considering actually a magnetic field.

The Dirac quantization for this constrained system has been carried out in Ref.\citen{rf:kokado2000}. 
First of all we diagonalize $F^\mu _{\  \nu }$ by matrices
\begin{align}
  & S^\mu{}_\nu
  = \frac{1}{\sqrt{2}}\begin{pmatrix} 1 &   1\\
                                     -i &   i  \end{pmatrix}~,
& & \left( S^{-1} \right)^\mu{}_\nu
  = \frac{1}{\sqrt{2}}\begin{pmatrix} 1 &   i \\
                                      1 &  -i \\ \end{pmatrix}~,
\label{eq:def-2S2}
\end{align}
to give
\begin{align}
  & \left( S^{-1} F S \right)^\mu{}_\nu
  = \begin{pmatrix} -i\,K& 0 \\
                       0 &  i\, K \end{pmatrix}~,
\label{eq:S_inv-2F2-S}
\end{align}
Then we define $X^{(\pm)}$ by
\begin{align}
  & S^{-1} \begin{pmatrix}  X^1 \\ X^2 \end{pmatrix}
  = \frac{1}{\sqrt{2}}\begin{pmatrix}  X^1 + i\,X^2  \\
                                       X^1 - i\,X^2  \end{pmatrix}
  \equiv  \begin{pmatrix} X^{(+)} \\ X^{(-)} \end{pmatrix}~.
\label{eq:def-2X2}
\end{align}
The boundary conditions (\ref{eq:bc_X}) become
\begin{align}
  & X'^{0}(\tau ,\sigma )
  \Big\vert_{\sigma=0, \pi} = 0~,
\label{eq:bc_X_0} \\
  & \big( X'^{(\pm )} \mp i\, \rho K \dot X^{(\pm )} \big)
  \Big\vert_{\sigma=0, \pi} = 0~.
\label{eq:bc_Xpm0}  
\end{align}
The solutions for Eqs.(\ref{eq:EOM_X}) and (\ref{eq:bc_Xpm0}) are
\begin{align}
  & X^{(\pm)}(\tau, \sigma)
  =  \sum_n\, \frac{ 1}{-i (n \pm \omega) }~e^{ -i (n \pm \omega) \tau}~
     \cos\big[~(n \pm \omega) \sigma \mp \pi \omega_0~\big]~ 
     \alpha^{(\pm)}_n + b^{(\pm)}~,
\label{eq:X-mode_decomposition}
\end{align}
where the cyclotron frequency $\omega $ is defined by
\begin{align}
\tan \pi \omega _0 \equiv \rho (0)K, \ \ 
\tan \pi \omega _\pi \equiv \rho (\pi )K, \ \\ 
\omega \equiv  \omega _0-\omega _\pi \quad \mbox{(we assume $\omega >0$)}.
\nonumber
\end{align}
The commutation relations for mode operators for $X^{(\pm)}$ are
\begin{align}
  & \big[\, \alpha^{(\pm)}_m\,,~\alpha^{(\mp)}_n~\big]
  = ( m \pm \omega  )~\delta_{m+n, 0}~,
& & \big[\, b^{(\pm)}\,,~b^{(\mp)}~\big]
  = - \frac{\cos \pi \omega _0 \cos \pi \omega _\pi }
           {\sin \pi \omega }~.
\label{eq:CCR_alpha}
\end{align}
We find the noncommutativity of $X^{(\pm)}(\tau, \sigma)$ at both ends, 
\begin{align}
  & \big[\, X^{(+)}(\tau, \sigma )\,,~X^{(-)}(\tau, \sigma' )~\big] = \left\{
   \begin{array}{rl}
    \theta_0~,\quad \sigma =\sigma '=0 \\
    \theta_\pi ~,\quad \sigma =\sigma '=\pi  \\
    0~, \quad \ \  \mbox{otherwise}
   \end{array}\right.
\label{eq:CCR_X-def_theta}
\end{align}
where
\begin{align}
   \theta_\sigma  \equiv  \frac{q_\sigma \, K}{ 1 + {q_\sigma }^2\, K^2 }~.
\label{eq:theta}
\end{align}

The solution for Eqs.(2.6) and (2.13) is the usual one, i.e.,
\begin{align}
 & X^0(\tau ,\sigma ) = x^0 + p^0\, \tau + i\sum_{n\neq 0}\frac{1}{n}e^{-in\tau }\alpha^0_n \cos (n\sigma )~,
\label{eq:mod_X0}
\end{align}
whereas for Eqs.(2.8) 
\begin{align}
 & X^a(\tau ,\sigma ) = c^a_0 + q^a\, \sigma + \sum_{n\neq 0}\frac{1}{n}e^{-in\tau }\alpha^a_n \sin (n\sigma )~,
\label{eq:mod_Xa}
\end{align}
where
\begin{align}
 & c^a_0 + q^a\, \pi = c^a_\pi ~.
\label{eq:relation_ca}
\end{align}
The commutation relations for mode operators are
\begin{align}
  & \big[\, \alpha ^{\mu }_m\,,~\alpha ^{\nu }_n~\big] = \eta ^{\mu \nu }\delta _{m+n,0}~,  \ \ \ \mu ,\nu =0, a~.
\label{eq:cc_al_al}
\end{align}

The Virasoro operator is defined as
\begin{align}
  & L_n
  = \frac{1}{4} \int^{\pi}_{-\pi} d\sigma~e^{\pm i n \sigma}~
      : \big( \dotX \pm X' \big)^2 :~
  = \half~\sum_{l}: \alpha_l \cdot \alpha_{n-l} :~,
\label{eq:def-Virasoro_op}
\end{align}
where $A\cdot B=-A^0B^0 + A^{(+)}B^{(-)} + A^{(-)}B^{(+)} +A^3B^3
+\cdot \cdot \cdot +A^{d-1}B^{d-1} $,
and satisfies the Virasoro algebra%
\footnote{The linear term in $\omega$ is necessary
in Eqs.(\ref{eq:def-Virasoro_algebra}) and \ref{eq:def-Virasoro_algebra2}).
The original article \cite{rf:kokado2000}
misses the term, so it should be modified in this way.}
\begin{align}
  & \big[\, L_m\,,~L_n~\big]
  = ( m - n )\, L_{m+n}
  + m\, \left\{~\frac{d}{12}\, (m^2 - 1) - \omega^2
     + \omega~\right\}~\delta_{m+n, 0}~,
\label{eq:def-Virasoro_algebra}
\end{align}
When $F^\mu _{\ \nu }$  has many blocks, Eq.(\ref{eq:def-Virasoro_algebra}) should be replaced by
\begin{align}
  & \big[\, L_m\,,~L_n~\big]
  = ( m - n )\, L_{m+n}
  + m\, \left\{~\frac{d}{12}\, (m^2 - 1) - \sum_i(\omega_i^{\ 2}
     - \omega_i~)\right\}~\delta_{m+n, 0}~,
\label{eq:def-Virasoro_algebra2}
\end{align}
where $\omega_i$ is the cyclotron frequency of the $i$-th block.

The Hamiltonian corresponds to $L_0$,
and the Virasoro conditions for physical states are
\begin{align}
  & L_n \ket{\psi} = 0~, \ \ n>0,
\nonumber \\
  & \left[ L_0 - \alpha (0) \right] \ket{\psi} = 0~.
\label{eq:physicalstate0}
\end{align}
where $\alpha(0)$ is Regge intercept.

\section{Spectcrum-generating algebra for the charged string}
\label{sec:SGAstring}
In this section we consider a SGA for charged strings 
when constant background fields are present. 
This is necessary to guarantee that the system is ghost-free.

First we recall SGA for a free string.
We define relevant operators as
\begin{align}
  & X^\mu (\tau )
  =  i \sum_{n \ne 0}\frac{1}{n}\, e^{ -i n \tau}~
    \alpha^{\mu }_n +x^\mu  + \tau p^\mu ~, 
\label{eq:X-mode_decomposition2} \\
  & P^\mu (\tau )
  = \partial _\tau X^\mu (\tau )
  = \sum_{n }\, e^{ -i n \tau}~\alpha^{\mu }_n~,
\label{eq:P-mode_decomposition} \\
  & V(\tau ) = e^{iX^{-}(\tau )}~,
\label{eq:V-def} \\
  & A^{i}_n = \oint d\tau ~P^i(\tau )V(\tau )^n~;
\quad
    \oint d\tau = (2\pi )^{-1}\int ^{\pi }_{-\pi }~d\tau~,
\quad i=1,\cdots ,d-2,
\label{eq:A-def} \\
  & K_n = - \oint d\tau :\{~P^{+}(\tau )
  + \frac{1}{2}n^2P^{-}(\tau )\log P^{-}(\tau )\}V(\tau )^n~:~,
\label{eq:K-def}
\end{align}
where $P^{\pm}$,$X^{\pm}$ are light-cone variables defined by
$X^{\pm}=(X^0 \pm X^{d-1})/\sqrt{2}$, etc., with 
the c.m. variables $x^\mu $, $p^\mu $,
and the mode operators $\alpha ^\mu _n$ satisfying
\begin{align}
  & \big[\, x^\mu \,,~p^\nu ~\big]=i\eta ^{\mu \nu}.
\label{eq:CCR_xp} \\
  & \big[\, \alpha^{\mu }_m\,,~\alpha^{\nu }_n~\big]
  = \eta ^{\mu \nu}m~\delta_{m+n, 0}~,
  \quad (\mu ,\nu =0,1,\cdots ,d-1)~.
\label{eq:CCR_alphamn}
\end{align}
We work in the subspace with $p^- = 1$, so that
\begin{align}
 & \oint d\tau ~P^-(\tau )V(\tau )^n=\delta _{n,0}~.
\label{eq:A-V}
\end{align}

SGA is summarized as follows:
\begin{align}
  & \big[\, A^i_m \,,~A^j_n ~\big]
  = m\delta  ^{ij}\delta _{m+n,0}~, \quad i,j=1,\cdots ,d-2
\nonumber \\
  & \big[\, A^{i }_m\,,~K_n~\big] = m~A_{m+n}^{i}~,
\label{eq:CCR_AmKn} \\
  & \big[\, K_m\,,~K_n~\big]
  = (m-n)~K_{m+n} + 2m^3\delta _{m+n,0}~.
\nonumber
\end{align}
The spectrum-generating operators $A^i_n$, $K_n$
are commutable with the Virasoro operator $L^{(0)}_n$
for the free string, and generate all the physical states
satisfying the Virasoro conditions.
The commutation relations (\ref{eq:CCR_AmKn}) suggest an isomorphism
\begin{align}
  & A^i_n \sim \alpha ^i_n~, \quad K_n \sim L^{(0) T}_n~,
\label{eq:isomorphism}
\end{align}
where $L^{(0) T}_n$ is the transverse part of the Virasoro operator
\begin{align}
   & L^{(0) T}_n \equiv  -a\delta _{n,0} + \frac{1}{2}\sum_{i=1}^{d-2}\,
   \sum_{l}\, :\alpha^i_{n-l} \alpha^i_{l}: ~, 
\label{eq:T-Virasoro-def} 
\end{align}
satisfying
\begin{align}
  & \big[\, \alpha ^i_m \,,~\alpha ^j_n ~\big]
  = m\delta  ^{ij}\delta _{m+n,0}~, \quad i,j=1,\cdots ,d-2
\nonumber \\
  & \big[\, \alpha ^{i }_m\,,~L^{(0) T}_n~\big]
  = m~\alpha ^{i }_{m+n}~,
\label{eq:CCR_AmKn-II} \\
  & \big[\, L^{(0) T}_m\,,~L^{(0) T}_n~\big]
  = (m-n)~L^{(0) T}_{m+n}
  + \big[\,\frac{d-2}{12}(m^3-m) + 2ma~\big]\delta _{m+n,0}~.
 \nonumber
\end{align}
The c-number $a$ is Regge intercept $a=\alpha (0)$.
Comparing Eqs.(\ref{eq:CCR_AmKn}) with (\ref{eq:CCR_AmKn-II}),
we find that the isomorphism is completed if $d=26$ and $a=1$.

Now let us return to the charged string
in Sec.\ref{sec:qunantize}. 
We first consider the case where $F^\mu _{\ \nu }$ has only one block. 
The components of string coordinates are given by 
$X^0, X^{(+)}, X^{(-)}, X^3, \cdots, X^{d-1}$.
Contrary to the free-end string, $X^a(\tau ,\sigma ), a=3, \cdots, d-1$, satisfy the fixed end boundary conditions.
 So, we should be careful for Eq.(3.1). Alternatively we define
\begin{align}
 & Y^{\mu }(s) = y^{\mu } + q^{\mu }s + i\, \sum_{n\neq 0} \frac{1}{n}\, e^{-ins}\alpha ^{\mu }_{n}~, \quad \mu =0,a~,
\label{eq:mod_Ymu}
\end{align}
where $y^{\mu }, q^{\nu }$ are a conjugate pair obeying $[y^{\mu }, q^{\nu }]=i\eta ^{\mu \nu }$, and
\begin{align}
  & P^{\mu } \equiv \partial _sY^{\mu }(s) = \sum \, e^{-ins}\, \alpha ^{\mu }_{n}, \ \ \mu =0, a~, \ \ \alpha ^{\mu }_0 = q^{\mu }~.
\label{eq:def_Pmu}
\end{align}
$X^0(\tau ,\sigma )$ is related with $Y^0$ as
\begin{align}
 & X^0(\tau ,\sigma ) = \frac{1}{2}\big[\, Y^0(\tau +\sigma ) + Y^0(\tau -\sigma ) ~\big] 
= y^0 + q^0\tau + i\, \sum_{n\neq 0}\, \frac{1}{n}\, e^{-in\tau }\, \alpha ^0_n \, \cos(n\sigma )~,
\label{eq:ret_X0}
\end{align}
so that $y^0=x^0, q^0=p^0$. On the other hand we have for $X^a(\tau , \sigma )$
\begin{align}
 & X^a(\tau ,\sigma ) = \frac{1}{2}\big[\, Y^a(\tau +\sigma ) - Y^a(\tau -\sigma ) ~\big] + c^a_0 
= c^a_0 + q^a\sigma  + \sum_{n\neq 0}\, \frac{1}{n}\, e^{-in\tau }\, \alpha ^a_n \, \sin(n\sigma )~,
\label{eq:ret_Xa}
\end{align}
where there appears no $y^a$ conjugate to $q^a=\alpha ^a_0$, which is the displacement operator of $\sigma $. \\
\indent We now consider the light-cone variables. The more general light-cone variables are defined by
\begin{align}
 & X^{\pm}(\tau ) \equiv  \frac{1}{\sqrt{2}}\kappa ^{\pm 1}\big[\, Y^0(\tau ) \pm Y^{d-1}(\tau ) ~\big]  
= x^{\pm} + p^{\pm}\tau  + i\, \sum_{n\neq 0}\, \frac{1}{n}\, e^{-in\tau }\, \alpha ^{\pm}_n~,
\label{eq:ret_Xpm} \\
 & P^{\pm}(\tau ) = \partial _{\tau } X^{\pm}(\tau ) = \sum \, e^{-in\tau }\, \alpha ^{\pm}_n~,
\label{eq:ret_Ppm}
\end{align}
where $\kappa $ is a real parameter, and
\begin{align}
 & x^{\pm} \equiv \frac{1}{\sqrt{2}}\kappa ^{\pm 1}(x^0 \pm y^{d-1})~, \ \ 
p^{\pm} \equiv \frac{1}{\sqrt{2}}\kappa ^{\pm 1}(p^0 \pm q^{d-1})~, 
\label{eq:def_xpm_ppm} \\
 & \alpha ^{\pm}_n \equiv \frac{1}{\sqrt{2}}\kappa ^{\pm 1}(\alpha ^0_n \pm \alpha ^{d-1}_n)~,
\nonumber \\
 & \big[\, p^{\pm}\,,~x^{\mp}~\big] = i~.
\label{eq:CCR_ppm_xmp}
\end{align}
As discussed in the last section, the usual light-cone variables with $\kappa =1$ are inadequate for the D-brane physics.

For the $(\pm)$ components(which should be distinguished from the light-cone variables $\pm$), instead of $P^\mu (\tau )$
in Eq.(\ref{eq:P-mode_decomposition}) we adopt
\begin{align}
  P^{(\pm)}(\tau )= \sum_{n}\,e^{-i(n\pm \omega )\tau }\alpha ^{(\pm)}_n~,
\label{eq:P_mode_exp}
\end{align}
which satisfy
\begin{align}
  & \big[\, P^{(\pm)}(\tau )\,,~P^{(\mp)}(\tau' )\big]
  = \sum_{n}\,(n \pm \omega )e^{-i(n\pm \omega )(\tau - \tau')}
  = i\partial _\tau \delta _{\pm\omega }(\tau - \tau')~.
\label{eq:CC_PP}
\end{align}
Here $\delta _{\omega }(s)$ is defined
as $\delta _{\omega }(s)=\exp(-i\omega s)\delta(s)$
with 2$\pi$-period delta function $\delta(s)$,
and has a property, $\delta _{\omega }(s+2\pi )
=\exp(-2\pi i\omega)\delta_{\omega }(s)$.

Corresponding to $A^{i}_n$ in Eq.(\ref{eq:A-def})
we define the SGA operators for the $(\pm)$ components as
\begin{align}
 & A^{(\pm )}_n=\oint d\tau ~P^{(\pm )}(\tau )V(\tau )^{n\pm \omega }~,
\label{eq:Apm=def}
\end{align}
where $V(\tau )$ is defined by $V(\tau ) = \exp[iX^-(\tau )]$ with the light-cone variable $X^-(\tau )$ above.
For the sake of $p^{-}=1$, $V(\tau )$ carries
a factor $e^{i\tau }$ and hence $V(\tau )^{n \pm \omega}$
a factor $e^{i(n\pm \omega )\tau }$.
The non-periodic factor $e^{i\omega \tau }$
is canceled by the similar factor in  $P^{(\pm)}(\tau)$.
In the following calculations this cancellation
always occurs among $\delta _{\pm \omega }, V(\tau )^{n \pm \omega}$
and $P^{(\pm)}$.
 
We are to show the commutation relations
\begin{align}
  & \big[\, A^{(\pm)}_m \,,~A^{(\mp)}_n ~\big]
  =(m \pm \omega )\delta _{m+n,0}~,
\label{eq:CCR_ApmAmp2} \\
  & \big[\, A^{(\pm )}_m\,,~K_n~\big] = (m\pm \omega )~A^{(\pm)}_{m+n}~.
\label{eq:CCR_AmKn-III}
\end{align}
which guarantee the $(\pm)$ parts of the isomorphism
$A^{(\pm)}_n \sim \alpha ^{(\pm)}_n$, and to show also that
the $A^{(\pm)}_n$'s commute with the Virasoro operators. 
Here $K_n$ is the same as Eq.(3.5), but the light-cone 
variables are defined by Eqs.(3.17) and (3.18).

The commutator  $\big[\, A^{(\pm)}_m \,,~A^{(\mp)}_n ~\big]$
is calculated as
\begin{align}
   \big[\, A^{(\pm)}_m \,,~A^{(\mp)}_n ~\big]
  &= \oint d\tau \oint d\tau '
     i\partial _\tau \delta _{\pm \omega }(\tau -\tau ')
     V(\tau )^{m\pm \omega }V(\tau ')^{n\mp \omega }
\nonumber \\
  &= \oint d\tau V(\tau )^{m\pm \omega }
  i\partial_\tau V(\tau )^{n\mp \omega }
  = -(n\mp \omega )\oint d\tau P^{-}(\tau )V(\tau )^{m+n}
\label{eq:CCR_ApmAmp3}
\\
  &= -(n \mp \omega )\delta _{m+n,0}~, \nonumber 
\end{align}
giving Eq.(\ref{eq:CCR_ApmAmp2}).
In the same way we have
\begin{align}
   \big[\, A^{(\pm )}_m\,,~K_n~\big]
  &= - \oint d\tau \oint d\tau ' P^{(\pm)}(\tau )
       \big[\, V(\tau )^{m\pm \omega }\,,~P^{+}(\tau ')~\big]
       V(\tau ')^{n}
\nonumber \\
  &= \oint d\tau \oint d\tau ' P^{(\pm)}(\tau )
     V(\tau )^{m\pm \omega }(m\pm\omega )
     \delta (\tau -\tau ')V(\tau ')^{n}
\nonumber \\
  &= (m\pm\omega )\oint d\tau P^{(\pm)}(\tau )
     V(\tau )^{m + n \pm \omega }~,
\label{eq:CCR_Am+-Kn}
\end{align}
to show Eq.(\ref{eq:CCR_AmKn-III}).

The commutativity of $A^{(\pm)}_n$ with the Virasoro operator
$L_n$ (\ref{eq:def-Virasoro_op}) with $F \neq 0$
can be shown as follows: We define $L(\tau)$ by
\begin{align}
  & L(\tau ) = \sum_{n}e^{-in\tau }L_{n}
  =-: P^{+}(\tau )P^{-}(\tau ): + :P^{(+)}(\tau )P^{(-)}(\tau ):
  + \frac{1}{2}\sum_{i=3}^{d-2}\,:\{P^i(\tau )\}^{2}: ~.
\label{eq:L-exp}
\end{align}
In order to show  $\big[\, L(\tau ) \,,~A^{(\pm)}_n ~\big]=0$,
we first calculate the first term of $L(\tau )$  as
\begin{align}
   \big[\, :P^{+}(\tau )P^{-}(\tau ):\,,~A^{(\pm)}_n ~\big]
  &= \oint d\tau' P^{-}(\tau )
     \big[\, P^{+}(\tau)\,,~V(\tau ')^{n\pm \omega }~\big]
     P^{(\pm )}(\tau ')
\nonumber \\
  &= \oint d\tau ' P^{-}(\tau )(n\pm \omega )
     \delta (\tau - \tau')V(\tau ')^{n\pm \omega }
     P^{(\pm)}(\tau ')
\nonumber \\
  &= -(n \pm \omega )P^{(\pm)}(\tau )P^{-}(\tau )
     V(\tau )^{n\pm \omega }~.
\nonumber
\end{align}
For the second term we have
\begin{align}
   \big[\, :P^{(+)}(\tau )P^{(-)}(\tau ):\,,~A^{(\pm)}_n~\big]
  &= \oint d\tau ' P^{(\pm)}(\tau )\big[\, P^{(\mp)}(\tau)\,,~
     P^{(\pm)}(\tau ')~\big]V(\tau ')^{n\pm \omega }
\nonumber \\
  &= \oint d\tau ' P^{(\pm)}(\tau )i\partial_{\tau }
     \delta _{\mp\omega }(\tau - \tau')V(\tau ')^{n\pm \omega }
\nonumber \\
  &= P^{(\pm)}(\tau )i\partial _{\tau }V(\tau )^{n\pm \omega }
  = -(n\pm\omega )P^{(\pm)}(\tau )P^{-}(\tau )V(\tau )^{n\pm\omega }~.
\nonumber
\end{align}
Since the third term is irrelevant, we get
\begin{align}
  & \big[\, L(\tau ) \,,~A^{(\pm)}_n ~\big]=0~.
\label{eq:CCL_Apmn}
\end{align}

To sum up, the modified SGA for the $(\pm)$-components is given by
\begin{align}
  & \big[\, A^{(\pm)}_m \,,~A^{(\mp)}_n ~\big]
  = (m \pm \omega )\delta _{m+n,0}~, 
\nonumber \\
  & \big[\, A^{(\pm)}_m\,,~K_n~\big]
  = (m \pm \omega )~A^{(\pm)}_{m+n}~,
\label{eq:CCR_AmKn-IV} \\
  & \big[\, K_m\,,~K_n~\big]
  = (m-n)~K_{m+n} + 2m^3\delta _{m+n,0}~.
\nonumber
\end{align}
These are to be compared with the commutation relations
\begin{align}
  & \big[\, \alpha ^{(\pm)}_m \,,~\alpha ^{(\mp)}_n ~\big]
  = (m \pm \omega )\delta _{m+n,0}~, 
\nonumber \\
  & \big[\, \alpha ^{(\pm)}_m\,,~L_n^T~\big]
  = (m \pm \omega )~\alpha ^{(\pm)}_{m+n}~,
\label{eq:CCR_alphamLn} \\
  & \big[\, L_m^T\,,~L_n^T~\big]
  = (m-n)~L_{m+n}^T + [\frac{d-2}{12}(m^3-m)
  - m(\omega ^2-\omega -2a)]\delta _{m+n,0}~.
\nonumber
\end{align}
where
\begin{align}
  & L_{n}^{T} \equiv  -a\delta _{n,0}
  + \frac{1}{2}\sum_{l}:[\alpha ^{(+)}_{n-l}\alpha ^{(-)}_{l}
  + \alpha ^{(-)}_{n-l}\alpha^{(+)}_{l} + \sum_{i=3}^{d-2}~\alpha ^{i}_{n-l}\alpha ^{i}_{l}]:~.
\label{eq:L-exp-II}
\end{align}
Comparing Eqs.(\ref{eq:CCR_AmKn-IV}) with (\ref{eq:CCR_alphamLn}),
the isomorphism
\begin{align}
  & A_n^{(\pm)} \sim \alpha _n^{(\pm)}~, \quad K_n \sim L_n^{T},
\label{eq:isomorphism2}
\end{align}
is completed if
\begin{align}
   d=26~, \quad
   a \equiv \alpha (0)=1-\frac{\omega }{2} + \frac{\omega ^2}{2}~.
\label{eq:d_alpha-condition}
\end{align}
When $F^\mu_{\ \nu }$ has many blocks, the last equation of Eqs.(\ref{eq:CCR_alphamLn}) should be replaced by
\begin{align}
  & \big[\, L_m^T\,,~L_n^T~\big]
  = (m-n)~L_{m+n}^T + [\frac{d-2}{12}(m^3-m)
  - m(\sum_i {\omega_i }^2-\sum_i \omega_i -2a)]\delta _{m+n,0}~.
\label{eq:CCR_LT_LT}
\end{align}
and the conditions (\ref{eq:d_alpha-condition}) by
\begin{align}
   d=26~, \quad
   a \equiv \alpha (0)=1-\frac{1}{2}\sum_i\omega_i  + \frac{1}{2}\sum_i {\omega_i}^2~.
\label{eq:d_alpha-conditionF}
\end{align}
where $\omega _i$ is the cyclotron frequency of the $i$-th block.

The other transverse components obey the same algebra
as in the free case, and we have the same isomorphism
as Eqs.(\ref{eq:isomorphism}). 
SGA thus obtained includes the operators $\alpha ^a_0 = q^a$. 
The $q^a$ should be replaced by 0-eigenvalue 
when both ends of the string end on the same brane, 
whereas by $q^a=(x^a_\pi  - x^a_0)/\pi $-eigenvalue when the other end-point 
of the open string is attached to the different brane.
Once we have SGA together with the Virasoro condition(\ref{eq:physicalstate0}), 
then we can conclude that our system is ghost-free provided the conditions (\ref{eq:d_alpha-conditionF})
are satisfied.

\section{The Hall conductivity} \label{sec:Hall_cond}

As an application, we consider the Hall effect for charged strings on the D2-brane, 
in which constant electromagnetic fields are living. We calculate the Hall conductivity 
for the current as a flow of charged strings.

The Hall effect is characterized by the induced electric field $E_1$ in the 1-direction 
on the plate (D2-brane), and the induced current density $J_2=\sigma _{21}E_1$ in the 2-direction, 
where $\sigma _{21}$ is the Hall conductivity, while there is no induced current in the 1-direction, 
i.e., $J_1=0$.

For this purpose we will concentrate on a $3\times 3$ block of $F^\mu _{\ \nu }$,
\begin{align}
  & F^{\mu}_{\ \nu}
  = \begin{pmatrix} 0 &  E & 0 \\ E & 0 & B \\ 0 & - B & 0
    \end{pmatrix}~.
\label{eq:def-F}
\end{align}
This is diagonalized as
\begin{align}
  & \left( S^{-1} F S \right)^\mu{}_\nu
  = \begin{pmatrix} 0 &      0 & 0 \\
                    0 & -i\, K & 0 \\
                    0 &      0 &  i\, K \end{pmatrix}~,
\label{eq:S_inv-F-S}
\end{align}
where
\begin{align}
  & S^\mu{}_\nu
  = \frac{1}{\sqrt{2}}\begin{pmatrix}
            a\sqrt{2} & i\, b & -i\, b \\
                    0 &   1   &  1 \\
          - b\sqrt{2} & -i\, a & i\, a \end{pmatrix}~,
& & \left( S^{-1} \right)^\mu{}_\nu
  = \frac{1}{\sqrt{2}}\begin{pmatrix} 
       a\sqrt{2} &    0 &   b\sqrt{2} \\
           i\, b &    1 & i\, a \\
          -i\, b &    1 & -i\, a \end{pmatrix}~,
\label{eq:def-S}
\end{align}
with $K = \sqrt{ B^2 - E^2 }$~(we assume $0 < E < B$) and
\begin{align}
  & a = \frac{B}{K}~, & & b = \frac{E}{K}~, & & a^2 - b^2 = 1~.
\end{align}
Then we have
\begin{align}
  & S^{-1} \begin{pmatrix} X^0 \\ X^1 \\ X^2 \end{pmatrix}
  = \begin{pmatrix} a\, X^0 + b\, X^2 \\
     \big( i\, b\, X^0 + X^1 + i\, a\, X^2 \big)/\sqrt{2} \\
     \big( -i\, b\, X^0 + X^1 - i\, a\, X^2 \big)/\sqrt{2} \end{pmatrix}
  \equiv  \begin{pmatrix} Z \\ X^{(+)} \\ X^{(-)} \end{pmatrix}~,
\label{eq:def-ZX}  
\end{align}
\begin{align}
  & \begin{pmatrix} X^0 \\ X^1 \\ X^2 \end{pmatrix}
  = S \begin{pmatrix} Z \\ X^{(+)} \\ X^{(-)} \end{pmatrix}
  = \begin{pmatrix} a\, Z +i\, b \big( X^{(+)} - X^{(-)} \big)/\sqrt{2} \\
                              \big( X^{(+)} + X^{(-)} \big)/\sqrt{2} \\
                  - b\, Z -i\, a\, \big( X^{(+)} - X^{(-)} \big)/\sqrt{2}
    \end{pmatrix}~.
\label{eq:rel-ZX}
\end{align}
The boundary condition Eq.(\ref{eq:bc_X}) become
\begin{align}
  & Z'(\tau ,\sigma )\, \Big\vert_{\sigma=0, \pi} = 0~, 
\label{eq:Zbc} \\
  & \big( X'^{(\pm )} \mp i\, \rho K \dot X^{(\pm )} \big)
  \Big\vert_{\sigma=0, \pi} = 0~.
\label{eq:XpmX'} 
\end{align}
So, $Z(\tau ,\sigma )$ plays a role of the time component of the free string. 
Its normal mode expansion is
\begin{align}
  & Z(\tau ,\sigma ) = z + p^0\tau + i\, \sum_{n \neq 0}\frac{1}{n} e^{-in\tau } \cos (n\sigma ) \alpha ^0_n~.
\label{eq:Zmodeex}
\end{align}
The other components, $X^3,\cdots, X^{d-1}$, obey the fixed end boundary conditions, 
and are already discussed before. The corresponding SGA is the same as for  
the case of $i=1$ in Sec.3.  

From Eq.(\ref{eq:rel-ZX}), $X^0$ and $X^2$ are given by
\begin{align}
  & X^0(\tau, \rho ) = a\, Z(\tau, \sigma ) + b\, Y(\tau, \sigma )~,
\label{eq:X_zero}
\\
  & X^2(\tau, \sigma ) = -b\, Z(\tau, \sigma ) - a\, Y(\tau, \sigma )
                 = -\frac{b}{a}\, X^0(\tau, \sigma ) - \frac{1}{a}\, Y(\tau, \sigma )~, 
\label{eq:x^2}
\end{align}
where
\begin{align}
 & Y \equiv  i\, ( X^{(+)} - X^{(-)} )/\sqrt{2}.
\label{eq:Y-def}
\end{align}

Now, let us define the ``time'' by an expectation value of $X^0$ 
\begin{align}
 & t \equiv \Exp{X^0(\tau ,0)} = \Exp{X^0(\tau ',\pi )} 
\nonumber \\
 &   =a(z + p^0\tau ) + b\Exp{Y(\tau ,0)} = a(z + p^0\tau ') + b\Exp{Y(\tau ',\pi )} ~.
\label{eq:def_t}
\end{align} 
Here the expectation values of physical quantities have been taken over 
stationary states, which are constructed by means of spectrum-generating operators obeying the 
new SGA, and, therefore, ghost-free. 
The $\tau $ derivative of Eq.(\ref{eq:def_t}) becomes
\begin{align}
 & \frac{\partial t}{\partial \tau } = ap^0 + b\Exp{\dotY(\tau ,0)} = ap^0=\frac{\partial t}{\partial \tau '}~,
\label{eq:dt_dtau}
\end{align}
because $\Exp{\dotY(\tau ,0)}$ vanishes, $\dotY \sim \dotX^{(\pm)}$ being linear in $\alpha ^{(\pm)}_n$.

The current density in the 2-direction is defined by
\begin{align}
 &  J^2 = q_0 n\frac{\partial }{\partial t}\Exp{X^2(\tau ,0)} + q_\pi n \frac{\partial }{\partial t}\Exp{X^2(\tau ',\pi )} 
\nonumber \\
 &      = - q_0 n \left[\frac{b}{a} + \frac{1}{a}\frac{\partial \tau }{\partial t}\Exp{\dotY(\tau ,0)}\right]
          - q_\pi  n \left[\frac{b}{a} + \frac{1}{a}\frac{\partial \tau '}{\partial t}\Exp{\dotY(\tau ',\pi )}\right] 
\label{eq:J2} \\
 &      = -(q_0 + q_\pi )n \frac{b}{a} =-(q_0 + q_\pi )n\frac{E}{B}~,
\nonumber 
\end{align}
where $n$ is the density of charged strings. Also we have
\begin{align}
 & \frac{\partial }{\partial t}\Exp{X^1(\tau ,0/\pi )}=\frac{\partial \tau }{\partial t}\frac{\partial }{\partial \tau }\Exp{X^1(\tau ,0/\pi )}
   =\frac{1}{ap^0}\Exp{\dotX^1(\tau ,0/\pi )}=0~,
\label{eq:J1}
\end{align}
because $\dotX^1=(\dotX^{(+)} + \dotX^{(-)})/\sqrt{2}$ is linear in $\alpha _n^{(\pm)}$. 
This means that the current density in the 1-direction vanishes, $J_1=0$.

In conclusion, the Hall conductivity for charged strings is given by $\sigma _{21}=-(q_0+q_\pi )n/B$, 
and this coincides with the same formula as in the ordinary particle Hall effect.
%
\section{Concluding remarks}
We have constructed the new SGA for charged strings placed in constant background fields. 
Contrary to the neutral string this algebra includes the cyclotron frequency. 
As a conclusion our system is ghost-free, if the space-time dimension is $d=26$ 
and the Regge intercept is  $\alpha (0)=1-\sum_i{\omega _i}/2+\sum_i{{\omega _i}^2}/2$, 
where $\omega _i$ is the $i$-th block cyclotron frequency of the charged string. 

In Eqs.(\ref{eq:ret_Xpm})-(\ref{eq:def_xpm_ppm}) we have defined the generalized light-cone variables, 
which involve the parameter $\kappa $. This parameter plays an important role in the D-brane physics. 
We had worked in the subspace $p^- = \kappa ^{-1}(p^0-q^{d-1})/\sqrt{2}=1$. On the other hand we have
\begin{align}
 & p^+ A^i_{-n} = A^i_{-n}\, (p^+ + n), \quad n=0,1,\cdots~.
\label{eq:P+A_n}
\end{align}
By operating the vacuum state from the r.h.s., we find $p^+ = \kappa (p^0+q^{d-1})/\sqrt{2}=n$. 
From both equations, $p^{-}=1$ and $p^+ = n$, we get non-negative solutions
\begin{align}
 & \kappa  = \frac{1}{\sqrt{2}}\Big( \sqrt{2n + {(q^{d-1})}^2} - q^{d-1} \Big), \quad 
  p^0 = \sqrt{2n + {(q^{d-1})}^2}~,
\label{eq:p0_qd-1}
\end{align}
where $q^{d-1}$ takes the zero value, i.e., $q^{d-1}=0$, when both ends of 
the open string end on the same D-brane, whereas $q^{d-1}$ takes the non-zero 
value when the other end-point of the string is attached to the different D-brane. 
Note that both equations, $p^- = 1$ and $p^+ = n$, give $p^0 \rightarrow q^{d-1}$ when $\kappa = n \rightarrow  0$. 
Thus we see that the usual light-cone variables with $\kappa =1$ are inadequate for the D-brane physics.

We have also calculated the Hall conductivity for the charged strings, 
which end on the D2-brane with a constant electromagnetic field. 
The result coincides with that of the ordinary two-dimensional electron system.

Finally we recall that, in order to obtain the quantum Hall effect for electrons, 
it is important to take into account of effects of impurity potentials 
or electron-electron scatterings. Otherwise, we get only the non-quantum result. 
The same thing may happen to charged strings. We expect the same quantum Hall effect for charged strings, 
if we take into account of effects of impurity potentials or string-string scatterings. 
These interesting problems will be remained in future studies.

\section*{Acknowledgements}
It is a pleasure to thank T. Okamura for continuous discussions. Thanks are also 
due to M. Kato for critical comments on D-brane boundary conditions.

\end{document}